\documentclass[conference,twocolumn,romanappendices]{IEEEtran}

\IEEEoverridecommandlockouts

\usepackage{algorithm,algorithmic,amsmath,amssymb,amsthm,bbm,caption,cite,color,comment,graphicx,microtype,setspace,subcaption,tabularx,url}
\usepackage[USenglish]{babel}
\usepackage[T1]{fontenc}
\usepackage[utf8]{inputenc}
\usepackage{hyperref}
\usepackage{csquotes}
\usepackage{mathtools}

\usepackage{enumitem}
\setitemize{itemsep=0mm,topsep=0mm,parsep=0pt,partopsep=0pt}

\usepackage{pgfplots}
\pgfplotsset{compat=newest}

\usepackage{tikz}
\usetikzlibrary{arrows,backgrounds,calc,intersections,decorations.pathreplacing,positioning,shapes}

\makeatletter
\newcommand\subparagraph{%
  \@startsection{subparagraph}{5}
  {\parindent}
  {3.25ex \@plus 1ex \@minus .2ex}
  {-1em}
  {\normalfont\normalsize\bfseries}}
\makeatother
\usepackage{titlesec}
\let\subparagraph\relax

\usepackage{titlesec}
\let\subparagraph\relax
\titlespacing{\section}{0pt}{5pt plus 2pt minus 1pt}{3pt plus 1pt minus 0pt}
\titlespacing{\subsection}{0pt}{4pt plus 2pt minus 1pt}{2pt plus 1pt minus 0pt}
\usepackage[figurename=Fig.,font=small]{caption}

{}
{}
{}
{\newtheorem{proposition}{Proposition}}
{\newtheorem{remark}{Remark}}
{}
{\newtheorem{corollary}{Corollary}}

\def\setS{\ensuremath{\mathcal{S}}}
\def\setS{\ensuremath{\mathcal{S}}}
\def\setC{\ensuremath{\mathcal{C}}}

\def\setN{\ensuremath{\mathcal{N}}}
\def\setU{\ensuremath{\mathcal{U}}}

\def\a{\ensuremath{\boldsymbol{a}}}

\def\n{\ensuremath{\mathbf{n}}}

\def\r{\ensuremath{\mathbf{r}}}

\def\y{\ensuremath{\mathbf{y}}}

\def\h{\ensuremath{\mathbf{h}}}

\def\I{\ensuremath{\mathbf{I}}}

\def\1{\ensuremath{\mathbf{1}}}
\def\0{\ensuremath{\mathbf{0}}}

\def\herm{\ensuremath{\text{H}}}
\def\tran{\ensuremath{\text{T}}}

\def\max{\ensuremath{\mathrm{max}}}

\newcommand{\re}{{\mathfrak{R}}}
\newcommand{\im}{{\mathfrak{I}}}

\newcommand\Prob[1]{\mathbb{P}\left\{#1\right\}}
\newcommand{\sgn}{\mathrm{sgn}}
\newcommand\qfunc[1]{Q\!\left(#1\right)}
\newcommand\gammaf[1]{\Gamma\left(#1\right)}

\newcommand\qonebit[1]{\ensuremath{\mathsf{Q}_{\mathrm{1b}}\left(#1\right)}}

\newcommand\br[1]{\left(#1\right)}

\newcommand\sqbr[1]{\left[#1\right]}

\newcommand{\Herm}{^{\mathrm{H}}}
\newcommand{\Trans}{^\mathrm{T}}

\DeclareMathOperator*{\argmax}{argmax}

\colorlet{blue}{black}
\colorlet{teal}{black}

\newcommand\smallO[1]{
    \mathchoice
    {% mode \displaystyle
      \scriptstyle\mathcal{O}{#1}
    }
    {% mode \textstyle
      \scriptstyle\mathcal{O}{#1}
    }
    {% mode \scriptstyle
      \scriptscriptstyle\mathcal{O}{#1}
    }
    {% mode \scriptscriptstyle
      \scalebox{0.8}{$\scriptscriptstyle\mathcal{O}$}{#1}
    }
}

\title{SEP Analysis  of $1$-Bit Quantized SIMO Systems with QPSK over Fading Channels}
\author{
\IEEEauthorblockN{Amila Ravinath, Minhua Ding, Bikshapathi Gouda, Italo Atzeni, and Antti Tölli}
\IEEEauthorblockA{Centre for Wireless Communications, University of Oulu, Finland \\
E-mail: \{amila.ravinath, minhua.ding, bikshapathi.gouda, italo.atzeni, antti.tolli\}@oulu.fi}
\thanks{This work was supported by the Research Council of Finland (336449 Profi6, 348396 HIGH-6G, 357504 EETCAMD, and 369116 6G~Flagship) and by the European Commission (101095759 Hexa-X-II).}}

\begin{document}

\maketitle

\begin{abstract}
The average symbol error probability (SEP) of  a $1$-bit quantized single-input multiple-output (SIMO) system  is analyzed under 
Rayleigh fading channels and quadrature phase-shift keying (QPSK) modulation. 
% The main challenge arises from the fact the conventional analytical methods relying on the decomposition of received signal and noise cannot be applied directly, as these two components are not well-defined after quantization.
%
Previous studies have partially characterized the diversity gain for selection combining (SC). In this paper, leveraging a novel analytical method,
%that overcomes the limitation of conventional approaches, 
an exact analytical SEP expression is derived for a $1$-bit quantized SIMO system employing QPSK modulation at the transmitter and maximum ratio combining (MRC) at the receiver. The corresponding diversity and coding gains of a SIMO-MRC system are also determined. Furthermore, the diversity and coding gains of a $1$-bit quantized SIMO-SC system are  quantified for an arbitrary number of receive antennas, thereby extending and complementing prior results.

\end{abstract}


%=========================================================================
\section{Introduction} \label{sec:INTRO}

Massive multiple-input multiple-output (MIMO) is one of the key enablers of current and next-generation wireless communication systems, offering significantly enhanced spectral efficiency and reliability~\cite{Rusek-et-al}. Employing a large number of antennas inevitably increases hardware cost,  complexity, and power consumption, especially for fully digital implementations.  It is well known that the power consumption of  analog-to-digital converters (ADCs) scales exponentially with the number of quantization bits~\cite{Emil-etal-2017, Lozano_2021}. This has motivated extensive research on low-resolution massive MIMO systems, and the use of $1$-bit quantization has received particular attention. A wide range of issues with $1$-bit quantized multi-antenna systems has been investigated, including capacity characterization and bounds~\cite{J_Mo_R_Heath, Lozano_2021}, 
channel estimation and data detection~\cite{Y_Li_et_al_BLMMSE, Atzeni_2022, Risi_Larsson_arxiv_paper, Choi_ML}, and symbol error probability (SEP) analysis~\cite{Gayan_2020_Open_J, Gay21}. 

The well-established results from using unquantized MIMO systems need to be re-examined, often with new approaches, when low-resolution quantization is applied~\cite{Lozano_2021}. For instance, while the capacities of single-input single-output (SISO) and multiple-input single-output (MISO) fading channels with $1$-bit ADCs at the receiver and perfect channel state information (CSI) at both the transmitter and the receiver were obtained in~\cite{J_Mo_R_Heath}, the capacities of  single-input multiple-output (SIMO) and MIMO fading channels remain unknown. Regarding the system reliability, the situation is parallel for the analysis of the average SEP (or simply SEP), a widely used performance index~\cite{Ber21, Ang_Li_et_al_2020, Wu_Liu_precoding_2024}. 
% Surprisingly, exact SEP results for fading multi-antenna systems with low-resolution ADCs are scarce.
In~\cite{Gayan_2020_Open_J}, the average SEP was studied for a low-resolution SISO system with $M$-ary phase-shift keying ($M$-PSK) modulation.  
In~\cite{Wu_Liu_et_al_2023}, by %upper- and lower-
bounding the SEP,  the diversity gain (or the diversity order)  of a MISO system was analyzed  with $M$-PSK modulation and low-resolution digital-to-analog converters (DACs) at the transmitter. Moreover, the impact of the number of quantization bits and the modulation order on the achievable diversity gain has been fully characterized in a MISO system~\cite{Wu_Liu_et_al_2023}. 

% Note that with
% perfect CSI at the transmitter, the MISO system model in~\cite{Wu_Liu_et_al_2023}
% reduces to an equivalent SISO model (cf.~\cite[Eq.~(3)]{Wu_Liu_et_al_2023}). 

In contrast, the SEP  of a SIMO system with $1$-bit ADCs
at the receiver has not been analyzed to the same {\color{black}extent}
%well investigated 
when complex-valued modulation schemes are used.
% due to the fact
% that there are no clearly defined received signal and noise
% components after quantization, thereby excluding the use of
% conventional analytical approaches.
The diversity gain of a quadrature phase-shift keying (QPSK)-modulated low-resolution SIMO system with selection combining (SC) at the receiver (SIMO-SC) was studied in~\cite{Gay21}, and was partially determined with $1$-bit quantization. 
% Clearly, the SEP analysis of SIMO systems is far from  complete. For instance, the diversity gain has not been obtained when the number of receive antennas is larger than two.
To delineate the SEP at high signal-to-noise ratio (SNR), in addition to the diversity gain, the coding gain
% , 
% i.e., {\color{black}the horizontal shift (in decibels) of the SEP curve in SNR relative to a benchmark curve}, 
is also an important parameter~\cite{Gia03,Ribeiro_Cai_Giannakis_2005}, which has not been discussed in the relevant literature. More importantly, although data detection based on maximum ratio combining (MRC) has
been widely used, e.g., in~\cite{Jac17, Y_Li_et_al_BLMMSE, Risi_Larsson_arxiv_paper,Atzeni_2022}, to the best of our knowledge, the exact SEP expression in a $1$-bit quantized SIMO system with MRC at the receiver (SIMO-MRC) has not been reported. 

% {\color{teal}For the diversity analysis, the primary tool we use at high SNR is \cite[Prop. 1]{Gia03} of which the intuition is that the SEP at high SNR is dominated by the low-probability event that the fading is small. Thus at high SNR it is enough to know the behavior of the probability density function (pdf) of the fading around $0$, the result of which is a bound for SEP at high SNR, parameterized by the diversity gain and the coding gain.}

In this work, we focus on the SEP analysis of a $1$-bit quantized SIMO system  {\color{black}under independent and identically distributed (i.i.d.) Rayleigh fading} with a large number of receive antennas and {\color{teal}with} QPSK modulation at the transmitter. {\color{black}Considering coherent detection and perfect CSI at the receiver (CSIR), the commonly used approach for analyzing the SEP of a SIMO-MRC system consists of two steps: first, the conditional SEP taking into account the effect of noise is obtained; then, the conditional SEP is averaged over the fading~\cite{Sim00, Gia03}. However, due to the quantization of the received signal, obtaining the conditional SEP given the CSIR is not straightforward, which further complicates the derivation of an average SEP expression.}  In view of the above, our contributions are summarized as follows:  i) 
% to circumvent the difficulty of applying conventional analytical methods, 
for a SIMO-MRC system, we  employ a new approach  that {\color{black}jointly} %utilizes the randomnesses of noise and fading and
{\color{black} leverages the \emph{circular symmetry} of} {\color{black}both the noise and fading distributions} to obtain an exact SEP expression for a $1$-bit quantized SIMO-MRC system, which allows to quantify the corresponding diversity and coding gains; ii) we complement the  study in~\cite{Gay21} by providing an alternative  diversity analysis of a $1$-bit quantized SIMO-SC system  and  deriving the corresponding coding gain.

\emph{Notation.} $a_n$ denotes the $n$th entry of the vector $\a$, $|a_n|$ the modulus of $a_n$, and  $\|\a\|_2$ the Euclidean norm of $\a$. $\a^*$ represents element-wise conjugation of $\a$.
$\a^\tran$ and $\a^\herm$ represent the transpose and Hermitian transpose of $\a$,
respectively. The  imaginary unit is denoted as $j=\sqrt{-1}$, $\0$ is the all-zero vector and $\I_n$ is the $n\times n$ identity matrix. $\setC\setN(\0,
\I_n)$ represents the zero-mean \emph{circularly symmetric} complex Gaussian distribution with
covariance matrix $\I_n$,  $\qfunc{x} =
\int_x^{\infty}\frac{e^{-t^2/2}}{\sqrt {2\pi}}{dt}$ is the Q-function,
%is the right tail probability
%of the standard normal distribution, 
and $\sgn(\cdot)$ stands for the signum
function.  $\re{(\cdot)}$ and $\im{(\cdot)}$ denote the real and imaginary parts,
respectively.  $\setU(a, b)$ denotes the uniform distribution over the interval $(a, b)$ and $\exp(\lambda)$ the exponential distribution with mean $\lambda$. $\mathbb{E}\{\cdot\}$ and $\mathbb{P}\{\cdot\}$ denote expectation and probability operators, respectively. $\smallO{(\cdot)}$ denotes the little-o notation. {\color{black} $x \to a^+$ denotes that $x$ tends to $a$ from above.} $\Gamma(z) = \int_0^\infty t^{z-1}e^{-t}\, {dt}$, with $\re\{z\}>0$, is the  Gamma function.
 
\section{System Model and Problem Statement} \label{sec:SM}
%=========================================================================
%\subsection{System Model}
Consider a SIMO system with ${N}$ receive antennas. The unquantized received  signal is denoted as
\begin{align}
\mathbf{y} = \sqrt{\rho}\mathbf{h}s +\mathbf{n}, \label{eqn: sys_model_original}
\end{align}
where the transmitted symbol $s$ is drawn uniformly from the QPSK constellation alphabet $\mathcal{S}=\{e^{j\br{\frac{\pi}{2}\left(i+\frac{1}{2}\right)}},\ i=0, 1,2,3\}$. The channel vector $\mathbf{h}=[h_1 \;\ldots \;h_{{N}}]\Trans$ and the noise vector $\mathbf{n}=[n_1\;\ldots \;n_{{N}}]\Trans$ are independent, with $\mathbf{h}, \mathbf{n}\sim\mathcal{CN}(\mathbf{0}, \mathbf{I}_{{N}})$. For clarity  in the subsequent discussions, let the entries of $\h$ be specified as
\begin{align}h_i=|h_i|e^{j\theta_i}, \label{eqn: fading_channel_components}
\end{align}
where $|h_i|^2\sim \exp(1)$ , $\theta_i\sim\mathcal{U}(-\pi, \pi)$,\ $i=1, \ldots, {N}$, and $|h_i|$ and $\theta_i$ are independent. Clearly, $\rho$ denotes the transmit SNR. Let 
\begin{align}
    \mathsf{Q}_{\rm 1b} (\cdot) =\frac{1}{\sqrt{2}}\br{\sgn(\re(\cdot)) + j\sgn(\im(\cdot))} \label{eqn: Q-1b_first_use}
\end{align}
represent the element-wise memoryless $1$-bit quantization operation.
Then, the quantized vector $\r$ at the output of the ADCs is   given by 
\begin{align}
    \mathbf{r} &= \mathsf{Q}_{\rm 1b} (\mathbf{y}) \in\setS^{N}.  \label{eqn: quantized observations}
\end{align}
Note that  $r_i\in\setS,\ i=1, \ldots, {N}$, due to \eqref{eqn: Q-1b_first_use}. 

%\subsection{Research Objectives}
We consider coherent detection by assuming perfect CSIR. Based on the quantized observations \eqref{eqn: quantized observations}, we investigate the average SEP performance of two diversity combining schemes, i.e., MRC and SC which are introduced in Section~\ref{sec: mrc} and Section~\ref{sec: sc}.

At high SNR, the average SEP $\mathbb{P}_{\rm SE}$ of a wireless system under fading can often be expressed as~\cite{Gia03, Proakis_Dig_comm} 
     \begin{align}
        \mathbb{P}_{\rm SE} \approx (G_{\rm c}\rho)^{-G_{\rm d}}, \quad\rho\rightarrow\text{large}. \label{eqn: high_SNR_SEP}
    \end{align}
Here, the two  parameters $G_{\rm d}$ and $G_{\rm c}$ are known as the diversity gain and the coding gain, respectively.   
% At high SNR, two important parameters -- the diversity  and coding gains, denoted as $G_{\rm d}$ and $G_{\rm c}$, respectively -- characterize the   average SEP, denoted as $\mathbb{P}_{\rm SE}$, of various systems under fading,  as shown below~\cite{Gia03}:
%      \begin{align}
%         \mathbb{P}_{\rm SE} \approx (G_{\rm c}\rho)^{-G_{\rm d}}, \quad\rho\rightarrow\text{large}. \label{eqn: high_SNR_SEP}
%     \end{align}
The importance and usage of these two parameters were exemplified, e.g., in~\cite{Gia03, Ribeiro_Cai_Giannakis_2005, Gayan_2020_Open_J, Gay21, Wu_Liu_et_al_2023}. {\color{black}To obtain $G_{\rm d}$ and $G_{\rm c}$, the primary tool we use is \cite[Prop. 1]{Gia03}, which facilitates efficient derivation of the above two gains once the behavior of the probability density function (pdf) of the relevant channel statistic is determined  around the origin. Different from those approaches used in~\cite{Wu_Liu_et_al_2023, Gayan_2020_Open_J, Gay21}, the one in \cite[Prop. 1]{Gia03} enables us to focus on extracting important channel statistics that determine the diversity and coding gains, as well as the essential calculations around it.}

For a SIMO-MRC system, we derive an exact  average SEP expression along with the corresponding diversity and coding gains. We also complement the result in~\cite[Thm.~4]{Gay21} by  obtaining the diversity and coding gains of the  SIMO-SC system for any value of $N$. While~\cite{Gayan_2020_Open_J,Gay21,Wu_Liu_et_al_2023} focused only on $G_{\rm d}$, here we quantify both  $G_{\rm d}$ and $G_{\rm c}$.

%=========================================================================
{\color{black}\section{Average SEP  of a $1$-Bit Quantized SIMO-MRC System with QPSK Modulation}\label{sec: mrc}}
Based on our system model, with perfect CSIR, MRC amounts to using the statistics of $\mathbf{h}\Herm\mathbf{r}$ for further processing~\cite{Jac17, Y_Li_et_al_BLMMSE, Risi_Larsson_arxiv_paper,Atzeni_2022}. 
Here, we focus on the following MRC-based QPSK symbol detection:
\begin{align}
    \widehat{s}_{\rm MRC}=\mathsf{Q}_{\rm 1b}(\mathbf{h}\Herm\mathbf{r}).\label{eqn: MRC_original}
\end{align} 
The corresponding average SEP is given by
\begin{align}
    \mathbb{P}_{\rm SE}^{\rm MRC}= \Prob{\widehat{s}_{\rm MRC}\neq s}, \label{eqn: MRC_SEP_definition}
\end{align}
where the randomnesses of $s$, $\h$, and $\n$ are considered.

{\color{blue}For subsequent use,  we introduce the half-normal pdf}
\begin{align}
    {\color{blue}f_1(v)=\sqrt{\frac{2}{\pi}}e^{-\frac{v^2}{2}}, \quad v>0. \label{eqn: pdf_1_sqrt}}
\end{align}
Our main result is given in the following proposition.

\begin{proposition}\label{prop: MRC_LS_MLD}
    The exact average SEP of a $1$-bit quantized QPSK-modulated SIMO-MRC system  with $N$ receive antennas under i.i.d. Rayleigh fading is expressed as 
    {\color{blue}
\begin{align}
    \mathbb{P}_{\rm SE}^{\rm MRC} &= 2 \mathbb{E}\left\{\qfunc{\sqrt{\frac{\rho}{N}U}}\right\} - \left(\mathbb{E}\left\{\qfunc{\sqrt{\frac{\rho}{N}U}}\right\}\right)^2
, \label{eqn: mrc_ser_NEW}
\end{align}
where $U=\left(\smash{\sum_{i=1}^NZ_i}\right)^2$ with  $\smash{\{Z_i\}_{i=1}^N}$ being i.i.d. half-normal random variables (cf. \eqref{eqn: pdf_1_sqrt}).}
The corresponding diversity gain $G_{\rm d, MRC}$ and coding gain  $G_{\rm c, MRC}$  are given by
    \begin{align}
         G_{\rm d, MRC} &= \frac{{N}}{2}, \label{eqn: MRC_diversity_gain}\\
        G_{\rm c, MRC} &= \left(\frac{2^{N}\pi^{-\frac{N+1}{2}}N^{\frac{N}{2}}}{{N}!}
        %{{N}!\sqrt \pi} \br{\frac {4 {N}}\pi}^{\frac {N} 2}
        %
        \Gamma\br{\frac{{N}+1}2}\right)^{-\frac2N},
        \label{eqn: MRC_high_SNR_coeffi}
    \end{align}
    respectively.
   \end{proposition}
\begin{IEEEproof}
Due to  space limitations, we provide only an outline of the proof. 
 {\color{black}More details can be found in~\cite{J}}. 

% Given perfect CSIR, the commonly used approach in multi-channel SEP analysis is a two-step process, starting with obtaining the conditional SEP which takes into account the effect of noise. The average SEP is further determined 
% by  averaging the conditional SEP  over the fading~\cite{Sim00, Gia03}.
{\color{black}As mentioned earlier, a conventional approach would start with the conditional SEP given perfect CSIR.} 
 For the MRC-based detection here, since $\mathbf{r}$ is the quantized output,  the decomposition of signal and noise component in  $\mathbf{h}\Herm\mathbf{r}$ is not well defined. As a result, it is not straightforward to obtain the conditional SEP given perfect CSIR, which further complicates the derivation of an average SEP expression. 
 
 Our new method here relies on dealing with the randomnesses of the channel and noise jointly. Capitalizing on the circular symmetry of the distributions of both $\h$ and $\n$, we establish that
\begin{equation}
\Prob{\qonebit{\h^\herm\qonebit{\y}} \ne s} = \Prob{\qonebit{(\qonebit{\h})^\herm\y} \ne s}. \label{eqn: key_identity}
\end{equation}
% {\color{black}A detailed proof of this identity is available in~\cite{J}.}
{\color{teal}
The proof of \eqref{eqn: key_identity} is based on the following three key facts: i) the joint distribution of the random
vector $\big[\h^\tran \ \frac{1}{\sqrt{\rho + 1}}\y^\tran\big]^\tran$ being identical to that of
$\big[\frac{1}{\sqrt{\rho + 1}}\y^\herm \ \h^\herm\big]^\tran$, ii) $\qonebit{kx} = \qonebit{x},\ k > 0$, and iii) $\qonebit{x^*} = \br{\qonebit{x}}^*$.}

The left-hand side of \eqref{eqn: key_identity} is simply \eqref{eqn: MRC_SEP_definition}. For a particular channel realization, on the right-hand side (RHS) of \eqref{eqn: key_identity}, the quantization operation is performed on the channel instead of on the received signal $\y$. Therefore,   $(\qonebit{\h})^\herm\y$  has a clear separation of signal and noise components, i.e.,
\begin{align}
(\qonebit{\h})\Herm\y=\sqrt{\rho}(\qonebit{\h})\Herm\h
s + (\qonebit{\h})\Herm\n, \label{eqn: equiv_model}
\end{align}
where
\begin{align}
(\qonebit{\h})\Herm\h&=\sum_{i=1}^N |h_i|e^{j\widetilde{\theta}_i},
\end{align}
   and $(\qonebit{\h})\Herm\n\sim\setC\setN(0,N)$.  {\color{blue} Here all $\widetilde{\theta}_i$'s are i.i.d., each with the distribution $\mathcal{U}\big(-\frac{\pi}{4}, \frac{\pi}{4}\big)$, whereas $|h_i|$ and $\widetilde{\theta}_i$ are independent, for all $i=1, \ldots, N$.
   
   Based on these facts, from the quantization associated with the RHS of \eqref{eqn: key_identity}, we obtain

   \begin{align}
    \mathbb{P}_{\rm SE}^{\rm MRC} &= 1 - \mathbb{E}\left\{Q\left(-\sqrt{\frac{\rho}{N}}\sum_{i=1}^{N}|h_i|\sqrt{1-\sin(2\widetilde{\theta}_i)}\right) \right.\nonumber
\\   &\phantom{=} \ \left.\times Q\left(-\sqrt{\frac{\rho}{N}}\sum_{i=1}^{N}|h_i|\sqrt{1+\sin(2\widetilde{\theta}_i)}\right)\right\}, \label{eqn: mrc_ser}
\end{align}
where the expectation is with respect to $|h_i|$ and $\widetilde{\theta}_i$, for all $i$. 
Furthermore, the following $2N$ random variables 
\[
|h_i|\sqrt{1-\sin(2\widetilde{\theta}_i)}, \quad |h_i|\sqrt{1+\sin(2\widetilde{\theta}_i)}, \quad i=1, \ldots, N\]
are shown to be i.i.d., each following the half-normal distribution in \eqref{eqn: pdf_1_sqrt}.
Consequently, the following two random variables
\[\br{\sum_{i=1}^{N}|h_i|\sqrt{1-\sin(2\widetilde{\theta}_i)}}^2, \quad\br{\sum_{i=1}^{N}|h_i|\sqrt{1+\sin(2\widetilde{\theta}_i)}}^2\]
are also i.i.d.  Then,  using the fact that $\qfunc{x}=1-\qfunc{-x}$,  we obtain \eqref{eqn: mrc_ser_NEW} from \eqref{eqn: mrc_ser},  where we have defined 
 \begin{align}U = \br{\sum_{i=1}^{N}|h_i|\sqrt{1-\sin(2\widetilde{\theta}_i)}}^2 \label{eqn: key_stat_MRC}
\end{align}
without loss of generality. Clearly, $U$ is the square of the sum of $N$ i.i.d. random variables, each with the pdf given in \eqref{eqn: pdf_1_sqrt}.  

At high SNR, the dominant term in \eqref{eqn: mrc_ser_NEW} is the first term on its RHS, where the key channel statistic  is  $U$ in \eqref{eqn: key_stat_MRC}.
The pdf of $U$} around the origin can be determined as
 \begin{align}\label{eqn: MRC_theorem_proof_4}
f_{U}(u) = c_1 u^{c_2} + {\smallO{\big(u^{c_2}\big)}}, \quad  u \to 0^+,
\end{align}
where 
\begin{align}
\label{eqn: MRC_theorem_proof_5}
c_1=\frac12\frac 1{({N}-1)!}\br{\frac2\pi}^{\frac {N}2}
, \quad c_2=\frac{{N}-2}2.
\end{align} 
Finally,  applying~\cite[Prop.~1]{Gia03}, 
we  obtain 
\eqref{eqn: MRC_diversity_gain} 
and \eqref{eqn: MRC_high_SNR_coeffi}.  
\end{IEEEproof}

\begin{figure}
\centering
\begin{tikzpicture} [auto]
\begin{semilogyaxis}
[
width=8cm,
height=6cm,
% put axis lines at left and bottom
% xmin=-30, xmax=40,
% ymin=5*10^-7, 
ymax=10^0,
ymin=10^(-8),
xmin=-10,
xmax=20,
ytick={10^0,10^-2,10^-4,10^-6, 10^-8},
% control axis labels
xlabel = {$\rho$ [dB]},
ylabel = {Average SEP},
ylabel near ticks,
x label style={font=\scriptsize},
y label style={font=\scriptsize},
ticklabel style={font=\scriptsize},
legend style = {font=\scriptsize, fill=white, fill opacity=0.6, text opacity=1},
legend pos = south west,
% control major grids
grid=both,
major grid style={line width=.2pt,draw=gray!30},
% every axis plot/.append style={thick},
mark options = {solid},
% legend to name={leg: mrc-mld}
% mark size = 1pt,
% cycle list name = exotic,
% cycle list shift = -4,
]

\addplot    [thick, dotted, mark=x, mark repeat=2] table [ y=mrc, x=rho, col sep=comma ] {data/mrcNr1.txt}; \addlegendentry{Simulated}
\addplot    [thick, dashed, mark=o, mark repeat=2] table [ y=exact, x=rho, col sep=comma ] {data/mrcNr1.txt}; \addlegendentry{Theoretical, {\color{blue}\eqref{eqn: mrc_ser_NEW} or \eqref{eqn: mrc_ser}}}
\addplot    [thick, solid]  table [ y=bound, x=rho, col sep=comma ] {data/mrcNr1.txt}; \addlegendentry{Bound \eqref{eqn: MRC_diversity_gain}, \eqref{eqn: MRC_high_SNR_coeffi}}

\addplot    [thick, dotted, mark=x, mark repeat=2] table [ y=mrc, x=rho, col sep=comma ] {data/mrcNr4.txt};
\addplot    [thick, dashed, mark=o, mark repeat=2] table [ y=exact, x=rho, col sep=comma ] {data/mrcNr4.txt};
\addplot    [thick, solid]  table [ y=bound, x=rho, col sep=comma ] {data/mrcNr4.txt};

\addplot    [thick, dotted, mark=x, mark repeat=2] table [ y=mrc, x=rho, col sep=comma ] {data/mrcNr8.txt};
\addplot    [thick, dashed, mark=o, mark repeat=2] table [ y=exact, x=rho, col sep=comma ] {data/mrcNr8.txt};
\addplot    [thick, solid]  table [ y=bound, x=rho, col sep=comma ] {data/mrcNr8.txt};

\addplot    [thick, dotted, mark=x, mark repeat=2] table [ y=mrc, x=rho, col sep=comma ] {data/mrcNr16.txt};
\addplot    [thick, dashed, mark=o, mark repeat=2] table [ y=exact, x=rho, col sep=comma ] {data/mrcNr16.txt};
% \addplot    [thick, solid]  table [ y=bound, x=rho, col sep=comma ] {data/mrcNr16.txt};
\addplot    [thick, solid, domain=0:12] {11.2263959688588 * 10^(-8*x/10)}; % Using formula: 2^N*pi^(-(N+1)/2)*N^(N/2)/factorial(N)*gamma((N+1)/2) for the high SNR coefficient in MATLAB
% \legend{Theoretical, Simulated}

\node[font=\footnotesize, anchor=north west] at (10.5, 9*10^-1) {$N=1$};
\node[font=\footnotesize, anchor=north west] at (10.5, 2.5*10^-2) {$N=4$};
\node[font=\footnotesize, anchor=north west] at (10.5, 3.5*10^-4) {$N=8$};
\node[font=\footnotesize, anchor=north west] at (10.5, 1*10^-7) {$N=16$};

\end{semilogyaxis}
\end{tikzpicture}
\caption{Average SEP vs. $\rho$ using MRC with $N\in\{1, 4, 8, 16\}$. The corresponding SEP bound is specified by \eqref{eqn: MRC_diversity_gain}--\eqref{eqn: MRC_high_SNR_coeffi}.}
\label{fig: mrc}
\end{figure}
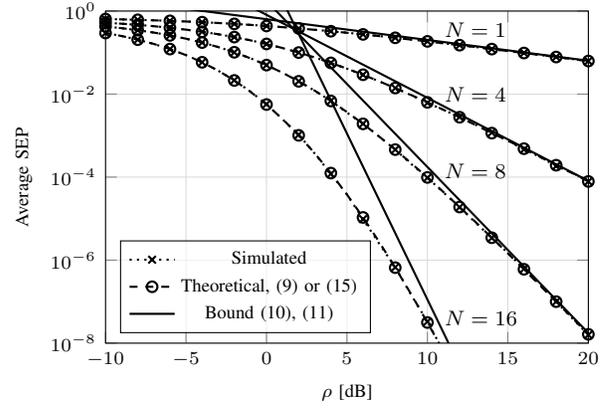

Fig.~\ref{fig: mrc} provides the SEP simulation results of a $1$-bit SIMO-MRC system for $N\in\{1, 4, 8, 16\}$, which corroborate the above SEP expression as well as its high-SNR characterization.

\begin{remark}
    It turns out that  \eqref{eqn: equiv_model} is precisely the same system model in the SEP analysis of a MISO system with quantized matched filter precoding at the transmitter~\cite[Eq.~(3)]{Wu_Liu_et_al_2023}. Therefore, \eqref{eqn: key_identity} reveals an interesting equivalence relationship between SIMO and MISO systems in terms of the average SEP with $1$-bit quantization and QPSK modulation under Rayleigh fading. A more general average SEP equivalence relation between SIMO and MISO systems with low-resolution quantization and $M$-PSK can also be established~\cite{J}.
\end{remark}

 The average SEP of an unquantized SIMO-MRC system with QPSK modulation under fading is given by~\cite[Eq.~(9.20)]{Sim00}
\begin{align}
    &\mathbb{P}_{\rm SE}^{\rm unq}
    =2\mathbb{E}\left\{Q\left(\sqrt{\rho\|\mathbf{h}\|_2^2}\right)\right\}-\mathbb{E}\left\{\left(Q\left(\sqrt{\rho\|\mathbf{h}\|_2^2}\right)\right)^2\right\}.\label{eqn: SEP_SIMO_unq}%\nonumber
\end{align}
The corresponding diversity gain from \eqref{eqn: SEP_SIMO_unq} under i.i.d. Rayleigh fading is $N$~\cite{Sim00, Gia03}. Comparing this with \eqref{eqn: MRC_diversity_gain}, it is clear that  $1$-bit quantization at the receiver incurs a loss of $N/2$ in the diversity gain.  
 Denote the coding gain from \eqref{eqn: SEP_SIMO_unq} under i.i.d. Rayleigh fading as $G_{{\rm c, MRC}}^{\rm unq}$, which is given by~\cite{Gia03}
 \begin{align}
 G_{\rm c, MRC}^{\rm unq} &= 2{{2N}\choose{N}}^{-\frac{1}{N}}.\label{eqn: MRC_high_SNR_coeffi_unq}
\end{align}
For the same diversity gain $N$, {\color{black} the coding gain of a $1$-bit quantized QPSK-modulated SIMO-MRC system with $2N$ receive antennas versus  that  of a corresponding unquantized system with $N$ receive antennas is given by:}  
% the coding gain of  $1$-bit quantized QPSK-modulated SIMO-MRC system with $2N$ receive antennas versus  that  of the corresponding unquantized system with $N$ receive antennas is given by
\begin{equation}
\begin{aligned}
\frac{G_{{\rm c, MRC}}(2N) }{G_{{\rm c, MRC}}^{\rm unq}(N) } &= \frac \pi{4 {N}}\left[\frac{\br{2N}!}{N!}\right]^{\frac1{N}} \in \left(\frac{\pi}{e}, \frac{\pi}{2}\right],\label{eqn: quant_unq_rat}
\end{aligned}
\end{equation}
where the lower bound is obtained by applying Stirling's formula to the factorials for large $N$ and the upper bound is achieved when $N=1$. For clarity, \eqref{eqn: quant_unq_rat} is also plotted in Fig.~\ref{fig: quant-unq-rat} in logarithmic scale versus $N$. 

Based on \eqref{eqn: quant_unq_rat}, we conclude that, to ensure the SEP achieved by an unquantized SIMO-MRC system at high SNR, it suffices to use twice the number of receive antennas when $1$-bit quantization is applied.

{\color{teal}
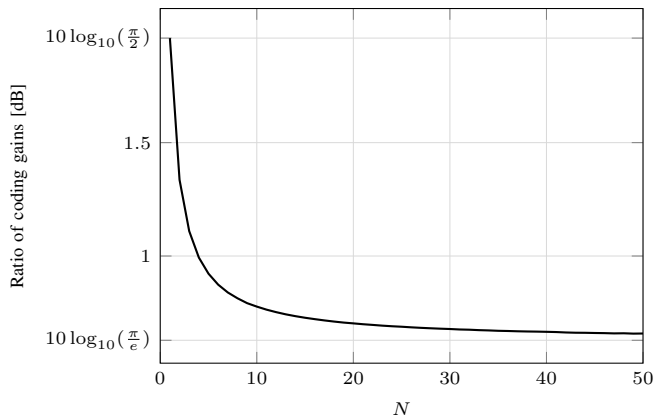
\begin{figure}
\centering
\begin{tikzpicture}[auto]
\begin{axis}[
width=8.0cm,
height=6.3cm,
xmin=0, xmax=50,
ylabel = {Ratio of coding gains [dB]},
xlabel = {$N$},
ylabel near ticks,
x label style={font=\scriptsize},
y label style={font=\scriptsize},
ticklabel style={font=\scriptsize},
ytick = {.5, 10*log10(pi/e), 1, 1.5, 10*log10(pi/2)},
yticklabels = {$0.5$, $10\log_{10}(\frac\pi e)$, $1$, $1.5$, $10\log_{10}(\frac \pi 2)$}, 
% control major grids
grid=both,
major grid style={line width=.2pt,draw=gray!30},
]
%Below the red parabola is defined
\addplot [
    domain=1:50, 
    samples=50, 
    thick,
]
{10*log10(pi/(4*x) * (factorial(2*x)/factorial(x))^(1/x))};% \addlegendentry{$\eqref{eqn: quant_unq_rat}$ [dB]}

% \addplot [
%     domain=1:50, 
%     samples=50, 
%     thick,
%     dotted,
% ]
% {10*log10(pi/exp(1))};% \addlegendentry{$10\log_{10}(\frac\pi e)$}
\end{axis}
\end{tikzpicture}
\caption{Ratio of the coding gains in \eqref{eqn: quant_unq_rat} vs. diversity gain $N$.}
\label{fig: quant-unq-rat}
\end{figure}}

% {\color{teal}For the large antenna array regime \eqref{eqn: mrc_ser} simplifies to $ \mathbb{P}_{\rm SE}^{\rm MRC} = 1-\br{\qfunc{-\sqrt{\frac{2}{\pi}\rho N}}}^2$ because $v_i = |h_i|\sqrt{1\pm \sin(2\theta_i)}$ is half-normally distributed so that $(\sum_{i = 1}^{N} v_i)/N \to \sqrt{\frac2\pi}$ when $N \to \infty$. }

\section{Average SEP  of a $1$-Bit Quantized SIMO-SC System with QPSK Modulation}\label{sec: sc}
In~\cite[Thm.~4]{Gay21}, the diversity gain of a $1$-bit quantized QPSK-modulated  SIMO system  was shown to be $\frac {N} 2$ with the maximum-distance selection scheme, but  only for ${N}\in\{1, 2\}$. 
% Clearly, the last piece of the puzzle in the diversity study of $1$-bit quantized SIMO-SC systems is yet to be solved. 
For completeness, here we provide an alternative perspective on~\cite[Sec.~IV-C]{Gay21} and  derive the selection diversity and coding gains for arbitrary $N$. 

{\color{blue}Recall the proof of Proposition~\ref{prop: MRC_LS_MLD}. From 
 \eqref{eqn: mrc_ser}, it is clear that, when $N=1$, the SEP of a SISO system is given by 
    \begin{align}
    \mathbb{P}_{{\rm SE, SISO}}&=\mathbb{E}\left\{\qfunc{\sqrt{\rho|h_1|^2(1-|\sin(2\widetilde{\theta}_1)|)}}\right\}
    \nonumber\\&\phantom{=} \
    +\mathbb{E}\left\{\qfunc{\sqrt{\rho|h_1|^2(1+|\sin(2\widetilde{\theta}_1)|)}}\right\}\nonumber
    \\&\phantom{=} \
    -\mathbb{E}\left\{\qfunc{\sqrt{\rho|h_1|^2(1-|\sin(2\widetilde{\theta}_1)|)}}\right.
    \nonumber
    \\&\phantom{=} \ \left.\times \qfunc{\sqrt{\rho|h_1|^2(1+|\sin(2\widetilde{\theta}_1)|)}}\right\},
    \label{eqn: SEP_M_eq_1}
\\&\to\frac{2}{\pi}\rho^{-1/2}, \quad \rho \to \text{large}, \label{eqn: asymptotic_SISO}   
\end{align}
where \eqref{eqn: asymptotic_SISO} is obtained from \eqref{eqn: MRC_diversity_gain}--\eqref{eqn: MRC_high_SNR_coeffi} with $N=1$. A form of \eqref{eqn: SEP_M_eq_1}--\eqref{eqn: asymptotic_SISO} for a SISO system was also obtained in~\cite[Eq.~(22)]{Gay17}. }
On the other hand,
the pdf of $|h_1|^2(1-|\sin(2\widetilde{\theta}_1)|)$  is given by
\begin{align}
f_2(v)=2\sqrt{\frac{2}{\pi v}}e^{-\frac{v}{2}}Q[\sqrt{v}], \quad v>0. \label{eqn: pdf_2}
\end{align} 
% which differs from that of $|h_1|^2(1\pm\sin(2\theta_1))$. 
From \eqref{eqn: pdf_2}, it can be shown that
\begin{align}
    &\mathbb{E}\left\{Q\left(\sqrt{\rho|h_1|^2(1-|\sin(2\widetilde{\theta}_1)|)}\right)\right\}\rightarrow \frac{2}{\pi}\rho^{-1/2},
    \label{eqn: discussions_SC}
\end{align} 
as $\rho\rightarrow\text{large}$. Comparing \eqref{eqn: asymptotic_SISO} and \eqref{eqn: discussions_SC}, we conclude that the first term on the RHS of   \eqref{eqn: SEP_M_eq_1} dominates $\mathbb{P}_{{\rm SE, SISO}}$ at high SNR, i.e.,
\begin{align}
    \mathbb{P}_{{\rm SE, SISO}}\rightarrow\mathbb{E}\left\{Q\left(\sqrt{\rho|h_1|^2(1-|\sin(2\widetilde{\theta}_1)|)}\right)\right\},  \label{eqn: discussion_SC_1}
    \end{align}
as $\rho\rightarrow\text{large}$.     
 
  Based on  \eqref{eqn: discussion_SC_1},     to minimize the SEP at high SNR, we select the antenna branch with the index obtained as
\begin{align}\argmax_{i\in\{1, \ldots, {N}\}}{|h_i|^2(1 -
|\sin2\widetilde{\theta}_i}|).\label{eqn: sel_criterion}
\end{align} 
Note that the above selection criterion %indicated by \eqref{eqn: sel_criterion} 
is equivalent to the maximum-distance selection in~\cite[Eq.~(13)]{Gay21} for a $1$-bit quantized QPSK-modulated SIMO-SC system. However, the simple form in \eqref{eqn: sel_criterion}  would facilitate further analyses. Once the branch is selected as in \eqref{eqn: sel_criterion}, the detected symbol is the one obtained based on the selected branch, and the resulting SEP is given by \eqref{eqn: SEP_M_eq_1} except that $h_1$ therein is replaced by the channel corresponding to the selected antennas. 

From  \eqref{eqn: discussion_SC_1} and \eqref{eqn: sel_criterion}, the asymptotic average SEP using SC is given by
\begin{align}
    &\mathbb{P}_{\rm SE}^{\rm SC}\rightarrow
\mathbb{E}%_{\mathbf{h}}
\left\{Q\left(\sqrt{\rho\underset{i\in\{1, \ldots, N\}}{\operatorname{max}}|h_i|^2(1-|\sin(2\widetilde{\theta}_i)|)}\right)\right\},  \label{eqn: discussion_SC_2}
\end{align}
as $\rho\rightarrow\text{large}$,
through which we can derive the corresponding diversity gain $G_{\rm d, SC}$ and coding gain  $G_{\rm c, SC}$  as in the following proposition.
\begin{proposition}
The diversity gain and coding gain of a $1$-bit quantized QPSK-modulated SIMO-SC system with the selection criterion \eqref{eqn: sel_criterion} under i.i.d. Rayleigh fading are given by
% >>>>>>> 33554c3290e8b33fab4b67efbc78263aab2c17aa
     \begin{align}
         G_{\rm d, SC} &= \frac{{N}}{2}, \label{eqn: Sel_diversity_gain}\\
        G_{\rm c, SC} &= \left(2^{2N-1}\pi^{-\frac{N+1}{2}}
        %\frac18\br{\frac{4}{\sqrt{\pi }}}^{{N}+1} 
        \gammaf{\frac{{N}+1}{2}}\right)^{-\frac{2}{N}},
        \label{eqn: Sel_high_SNR_coeffi}
    \end{align}
respectively.
\end{proposition}
\begin{IEEEproof}
    The analysis here is based on  \eqref{eqn: discussion_SC_2} with the following {\color{black}key statistic}: \[V_{\rm max} = \underset{i\in\{1, \ldots, N\}}{\operatorname{max}}|h_i|^2(1-|\sin(2\widetilde{\theta}_i)|).\] 
    {Clearly, $|h_i|^2(1-|\sin(2\widetilde{\theta}_i)|)$, $i=1, \ldots,N$, are i.i.d. with the pdf given by \eqref{eqn: pdf_2}, and their common  cumulative distribution function is given by
    \begin{align}
        F_2(v)=1 - 4\br{\qfunc{\sqrt{v}}}^2, \quad v>0.
    \end{align}
    } 
    Using order statistics~\cite{Papoulis_Pillai_Prob}, the pdf of $V_{\rm max}$  is
    \begin{align}
        f_{V_{\max}}(v) 
        &=N\sqbr{F_2(v)}^{N-1} f_2(v) \nonumber
        \\&= \frac{4{N}}{\sqrt{2\pi v}} e^{-\frac{v}{2}}\qfunc{\sqrt{v}}\sqbr{1 - 4\br{\qfunc{\sqrt{v}}}^2}^{{N}-1}  
    \end{align}
  for $v>0$. Around the origin, we have  
\begin{equation}\label{eq:pdf_T_max_approx}
\begin{aligned}
f_{V_{\max}}(v) &= d_1 v^{d_2} + {\smallO{(v^{d_2})}}, \quad v \to 0^+\\
\end{aligned}, 
\end{equation}
where
\begin{equation}\label{eqn: Sel_pdf_params}
d_1 = \frac{N}2\br{\frac{4}{\sqrt{2\pi }}}^{N}, \quad d_2 = \frac{N-2}{2}.\end{equation} Further applying~\cite[Prop.~1]{Gia03}, we obtain the desired results in \eqref{eqn: Sel_diversity_gain}--\eqref{eqn: Sel_high_SNR_coeffi}. 
%Details are omitted due to space limit.
\end{IEEEproof}   

% >>>>>>> 33554c3290e8b33fab4b67efbc78263aab2c17aa

\begin{figure}
\centering
\begin{tikzpicture} [auto]
\begin{semilogyaxis}
[
width=8cm,
height=6cm,
% put axis lines at left and bottom
xmin=10, xmax=30,
ymin=10^-8, ymax=10^0,
 ytick={10^0,10^-2,10^-4,10^-6,10^-8},
% control axis labels
xlabel = {$\rho$ [dB]},
ylabel = {Average SEP},
ylabel near ticks,
x label style={font=\scriptsize},
y label style={font=\scriptsize},
ticklabel style={font=\scriptsize},
legend style = {font=\scriptsize},
legend pos = south west,
% control major grids
grid=both,
major grid style={line width=.2pt,draw=gray!30},
% every axis plot/.append style={thick},
mark options = {solid},
% legend to name={leg: mrc-mld}
% mark size = 1pt,
% cycle list name = exotic,
% cycle list shift = -4,
]

\addplot    [thick, dotted, mark=x, mark repeat=2] table [ y=ser, x=rho, col sep=comma ] {data/selNr1.txt}; \addlegendentry{{\color{teal}Simulated}}
\addplot    [thick, dashed, mark=o, mark repeat=2] table [ y=serapprox, x=rho, col sep=comma ] {data/selNr1.txt}; \addlegendentry{{\color{teal}RHS of \eqref{eqn: discussion_SC_2}}}
\addplot    [thick, solid] table [ y=bound, x=rho, col sep=comma ] {data/selNr1.txt}; \addlegendentry{Bound \eqref{eqn: Sel_diversity_gain}--\eqref{eqn: Sel_high_SNR_coeffi}}

\addplot    [thick, dotted, mark=x, mark repeat=2] table [ y=ser, x=rho, col sep=comma ] {data/selNr2.txt};
\addplot    [thick, dashed, mark=o, mark repeat=2] table [ y=serapprox, x=rho, col sep=comma ] {data/selNr2.txt};
\addplot    [thick, solid] table [ y=bound, x=rho, col sep=comma ] {data/selNr2.txt};

\addplot    [thick, dotted, mark=x, mark repeat=2] table [ y=ser, x=rho, col sep=comma ] {data/selNr4.txt};
\addplot    [thick, dashed, mark=o, mark repeat=2] table [ y=serapprox, x=rho, col sep=comma ] {data/selNr4.txt};
\addplot    [thick, solid] table [ y=bound, x=rho, col sep=comma ] {data/selNr4.txt};

\addplot    [thick, dotted, mark=x, mark repeat=1] table [ y=ser, x=rho, col sep=comma ] {data/selNr8.txt};
\addplot    [thick, dashed, mark=o, mark repeat=1] table [ y=serapprox, x=rho, col sep=comma ] {data/selNr8.txt};
\addplot    [thick, solid] table [ y=bound, x=rho, col sep=comma ] {data/selNr8.txt};

\node[font=\footnotesize, anchor=north west] at (26, 1.5*10^-1) {$N=1$};
\node[font=\footnotesize, anchor=north west] at (26, 1.5*10^-2) {$N=2$};
\node[font=\footnotesize, anchor=north west] at (26, 2.5*10^-4) {$N=4$};
\node[font=\footnotesize, anchor=north west] at (26, 3*10^-7) {$N=8$};

\end{semilogyaxis}
\end{tikzpicture}
\caption{Average SEP vs.  $\rho$ using SC with $N\in\{1, 2, 4, 8\}$.  The corresponding SEP bound is specified by \eqref{eqn: Sel_diversity_gain}--\eqref{eqn: Sel_high_SNR_coeffi}.}
\label{fig: sc}
\end{figure}
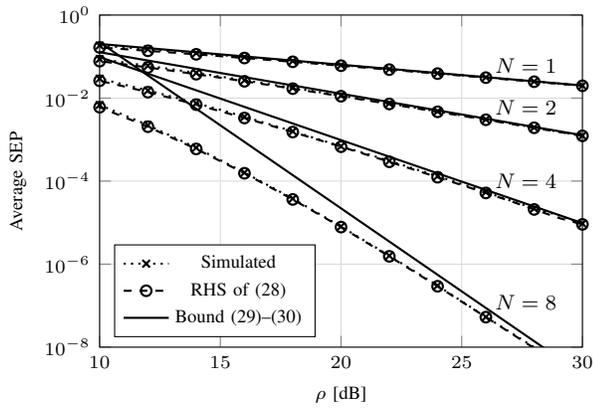

Fig.~\ref{fig: sc} depicts the high-SNR SEP performance parameterized using \eqref{eqn: Sel_diversity_gain}--\eqref{eqn: Sel_high_SNR_coeffi} together with the simulated average SEP of a $1$-bit quantized SIMO-SC system.

\begin{remark}
% We proved a conjecture in~\cite[Thm.~4]{Gay21} regarding the diversity gain of $1$-bit quantized QPSK modulated SIMO-SC systems under i.i.d. Rayleigh fading. 
 The MRC and SC with the selection criterion \eqref{eqn: sel_criterion} yield the same diversity gain for all values of $N$ and differ in coding gain for $N>1$, as in the unquantized case. When $N=1$,  both \eqref{eqn: Sel_diversity_gain}--\eqref{eqn: Sel_high_SNR_coeffi} and \eqref{eqn: MRC_diversity_gain}--\eqref{eqn: MRC_high_SNR_coeffi} agree with{\color{black}~\cite[Eq.~(22)]{Gay17}}. \end{remark}
 
 \begin{corollary}
    Given $N$ receive antennas in a $1$-bit quantized SIMO system,   the coding gain with MRC  versus  that  with SC is given by
\begin{equation}
\begin{aligned}
\frac{G_{{\rm c, MRC}}(N) }{G_{{\rm c, SC}}(N) } &= \frac{2^{2-\frac2N}}{N}\left(N!\right)^{\frac2{N}}. \label{eqn: MRC_Sel_rat}
\end{aligned}
\end{equation}
% {\color{teal}In \eqref{eqn: MRC_Sel_rat}, by applying Stirling’s formula, you could easily show that the ratio
% of the coding gains scales as $N^{1+1/N}$, that is, approximately linear. This
% is perhaps to be expected as in the unquantized case.}
 \end{corollary}

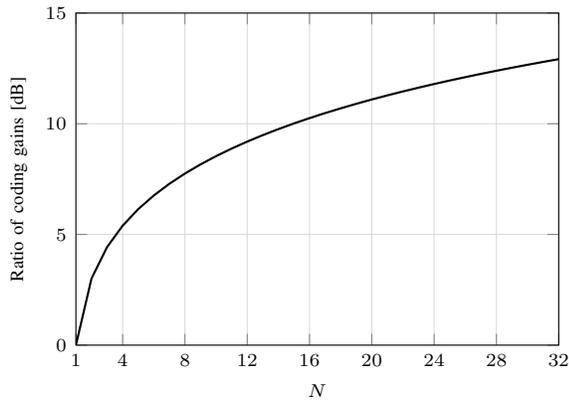
\begin{figure}
\centering
\begin{tikzpicture}[auto]
\begin{axis}[
width=8cm,
height=6cm,
xmin=1, xmax=32,
ymin=0, ymax=15,
xtick={1,4,8,...,32},
ylabel = {Ratio of coding gains [dB]},
xlabel = {$N$},
ylabel near ticks,
x label style={font=\scriptsize},
y label style={font=\scriptsize},
ticklabel style={font=\scriptsize},
% control major grids
grid=both,
major grid style={line width=.2pt,draw=gray!30},
]
%Below the red parabola is defined
\addplot [
    domain=1:32, 
    samples=32, 
    thick,
]
{10*log10(4/x * (factorial(x)/2)^(2/x))};
\end{axis}
\end{tikzpicture}
\caption{Ratio of coding gains of using MRC or SC in \eqref{eqn: MRC_Sel_rat} vs. number of receive antennas $N$.}
\label{fig: mrc-sel-rat}
\end{figure}

Fig.~\ref{fig: mrc-sel-rat} {\color{black} presents the ratio of coding gains in dB of the two diversity methods}  by plotting \eqref{eqn: MRC_Sel_rat} in dB. {\color{blue}For example, when $N=15$, there is a gain of approximately  $10$ dB  in transmit SNR from using MRC instead of SC.} The advantage of MRC over SC is shown to increase with $N$, as expected. 
 
%=========================================================================
\section{Conclusions} \label{sec:CONCL}
%=========================================================================

In this paper, we performed a comprehensive SEP analysis of a $1$-bit quantized QPSK-modulated SIMO system under i.i.d. Rayleigh fading. We derived an exact SEP expression for a $1$-bit quantized SIMO-MRC system. Two
high-SNR SEP parameters, i.e., the diversity
gain and coding gain, were quantified for both a $1$-bit quantized SIMO-MRC and a SIMO-SC system, which complemented previous diversity analysis of a SIMO-SC system in the literature. All the analyses were corroborated by simulations. {\color{blue}
Extension of this work, which takes into account 
higher-order phase modulation   and higher-resolution quantization, can be carried out along a similar line of development and is reported in~\cite{J}.}

\appendices

%=========================================================================
\bibliographystyle{IEEEtran}
\bibliography{IEEEabbr, conf_refs}
%=========================================================================

\end{document}